\documentclass[conference]{IEEEtran}
\usepackage{verbatim}
\usepackage{graphicx,epsfig}
\usepackage{amsfonts}
\usepackage{subfigure}
\usepackage[normalem]{ulem}
\usepackage{amsmath}

\def\bi{\begin{itemize}}
\def\ei{\end{itemize}}
\def\bequ{\begin{equation}}
\def\eequ{\end{equation}}

\def\benum{\begin{enumerate}}
\def\eenum{\end{enumerate}}

\begin{document}

\title{Cooperative Retransmissions Through Collisions}
\author{Jalaluddin Qureshi, Jianfei Cai and Chuan Heng Foh  \\
School of Computer Engineering\\
Nanyang Technological University, Singapore\\
\{jala0001, aschfoh, asjfcai\}@ntu.edu.sg}

\maketitle

\begin{abstract}
Interference in wireless networks is one of the key
capacity-limiting factors. Recently developed interference-embracing
techniques show promising performance on turning collisions into
useful transmissions. However, the interference-embracing techniques
are hard to apply in practical applications due to their strict
requirements. In this paper, we consider utilising the
interference-embracing techniques in a common scenario of two
interfering sender-receiver pairs. By employing opportunistic
listening and analog network coding (ANC), we show that compared to
traditional ARQ retransmission, a higher retransmission throughput
can be achieved by allowing two interfering senders to cooperatively
retransmit selected lost packets at the same time. This simultaneous
retransmission is facilitated by a simple handshaking procedure
without introducing additional overhead. Simulation results
demonstrate the superior performance of the proposed cooperative
retransmission.

\end{abstract}

\section{Introduction} \label{sect:Introduction}
Compared with centralized \emph{medium access control} (MAC)
protocols, random access based MAC protocols such as \emph{Carrier
Sense Multiple Access with Collision Avoidance} (CSMA/CA) do not
suffer from single point of failure and the network scalability
problem, and thus it has become dominant in \emph{wireless local
area networks} (WLANs). However, using random access for multiple
nodes to share a common channel inevitably introduces collisions or
interferences, especially at heavy traffic load.

Numerous approaches have been proposed to deal with wireless signal
interference. The common idea is to avoid collision as much as
possible. For example, CSMA/CA uses carrier sensing and random
backoff to avoid collision. Other techniques~\cite{Drieberg09}
include channel assignment, load balancing and power control. All
these techniques can alleviate wireless interference to a certain
extent, but cannot completely eliminate interference.

In 2006, Zhang \emph{et al.}~\cite{Zhang06} introduced a novel idea
of decoding a transmission collision on a wireless channel, which
directly challenges the traditional rule, that a collided
transmission on a wireless channel is undecodable. In this
pioneering work called \emph{physical-layer network coding}
(PNC)~\cite{Zhang06}, it shows that two simultaneous wireless
transmissions added together at the electromagnetic wave level can
be decoded to produce an outcome same as network coding. Katti
\emph{et al.}~\cite{Katti07} further elaborated this concept of
embracing wireless interference and proposed an \emph{analog network
coding} (ANC) scheme, which is more practical than PNC.

Despite this remarkable idea of turning collisions into useful
transmissions, it is hard to apply PNC and ANC in practical
applications. There are a few reasons for that. First, PNC and ANC
are only suitable for decoding a superimposed transmission
consisting only of two simultaneous transmissions. Second, the
collided transmissions need to be well synchronized, although a
perfect synchronization is not required for ANC. Third, in order
to decode a superimposed transmission, one of the two collided
transmissions needs to be known. All these constraints limit the
use of the interference-embracing techniques.

Although so far there is no practical solution to decode a
superimposition of multiple (more than two) transmissions, some
recently developed schemes~\cite{Foh10,Durvey07} show that it is
possible to tell the presence of individual transmissions involved
in a collision. These techniques have been successfully applied to
improve the reliability of wireless
broadcasting~\cite{Foh10,Durvey07}. The idea is that upon receiving
a broadcast transmission, each receiver detecting the transmission
replies with an \emph{acknowledge} (ACK) transmission. These
simultaneous ACK packets transmissions cause a collision. Then, decoding of
the superimposed ACK packet is performed to identify the ACK
transmitters. The synchronization issue is well handled in this case
since simultaneous ACK transmissions appear after the completion of
a broadcast transmission which is a common event.

\begin{figure}
\begin{center}
\includegraphics[width = 0.5\textwidth]{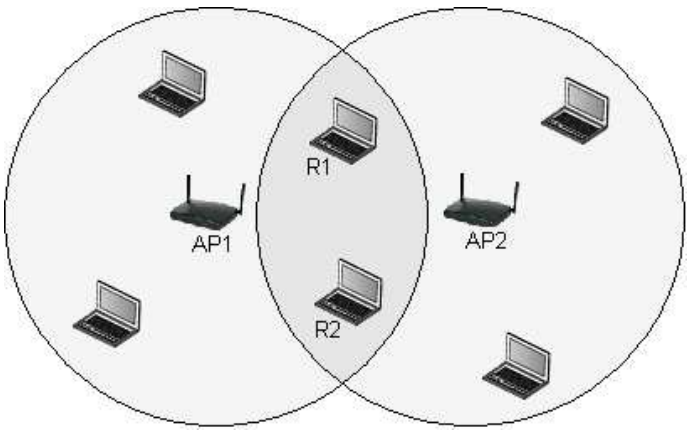}
\end{center}
\caption{Two interfering APs.} \label{fig:network}
\end{figure}

In this paper, we consider utilizing the interference-embracing
techniques in a common unit of two interfering sender-receiver
pairs. Particularly, we study the scenario of two interfering WLAN
APs, which are simulcasting bulky data to their associated individual
stations in a lossy environment as shown in Fig.~\ref{fig:network}.
This scenario is in line with the increasing density of WLAN APs and
the increasing popularity of multimedia applications such as video
streaming and online games~\cite{Drieberg09,Akella05}. Due to the
high performance to price ratio, more and more WLANs are being
deployed in public and residential places. Thus, it is quite common
that multiple APs overlap with each other and share a common
channel, especially in metropolitan cities.

By employing opportunistic listening and ANC, we show that compared
to traditional ARQ retransmission, a higher retransmission
throughput can be achieved by allowing two interfering APs to
cooperatively retransmit selected lost packets at the same time.
This simultaneous retransmission is facilitated by a simple
handshaking procedure without introducing additional overhead.
Simulation results demonstrate the superior performance of the
proposed cooperative retransmission.

The rest of the paper is organized as follows.
Section~\ref{sect:related} reviews the existing
interference-embracing techniques. Section~\ref{sect:proposed}
introduces our ideas and describes the detailed protocol design.
We analyze the proposed collision based retransmission scheme in
Section~\ref{sect:analysis} and provides the simulation results in
Section~\ref{sect:simu}. Finally, we conclude the paper in
Section~\ref{sect:conclude}.

\section{Related Work} \label{sect:related}
In this section, we review the existing interference-embracing
techniques, including interference based network coding and
superimposed acknowledgement. These techniques will be
incorporated into our proposed scheme as individual components.

\subsection{Interference Based Network Coding}
The idea of turning a collision of two simultaneous wireless
transmissions into a useful transmission was first introduced in
PNC~\cite{Zhang06}. In particular, the authors proposed a
frame-based decode-and-forward strategy in packet forwarding. In
their scenario of a relay network, two nodes transmit simultaneously
to a common receiver. Assuming perfect transmission synchronization
at the physical layer, based on the additive nature of
simultaneously arriving \emph{electromagnetic waves} (EM), the
receiver detects a single collided signal which is the sum of the
two transmitted signals. Using a suitable mapping scheme, they show
that for certain modulation schemes, there exists a mapping scheme such
that the relationship between the two transmitted binary bits and
the decoded binary bit follows the \emph{exclusive-or} (XOR)
principle.

ANC~\cite{Katti07} was further proposed to relax the restrictions of
symbol-level synchronization, carrier-frequency synchronization and
carrier-phase synchronization required in PNC, which makes ANC more practical.
Specifically, ANC is able to decode an unknown packet $c_2$ from a collided
packet $c_1\odot c_2$\footnote{ We use the notation $\odot$ to denote a
collision of two packet transmissions. } based on the known packet $c_1$ by
leveraging the co-channel FM signal separation technique~\cite{Hamkins00} and
network layer information to cancel the interference.

\subsection{Superimposed Acknowledgement}
As mentioned, some interesting methods~\cite{Foh10,Durvey07} have
been proposed to decode superimposed ACKs for providing reliability
in wireless multicast, where the requirement is not to decode the
content of the collided packets but to detect the existence of
individual ACKs from different receivers. In \cite{Durvey07}, Durvy
\emph{et al.} proposed to use a bit sequence of $N+1$ bits to decode
a collision of up to $N$ simultaneous ACK transmissions. The main
limitation of the scheme is that it requires precise power level
differentiation in the decoding procedure. Comparisons of analog
received signals are needed for the operation, and a delay line is
used to store analog signals for the comparison purpose.

In our previous work~\cite{Foh10}, we design a coding method, called
collision codes, used in the MAC layer that can also achieve the
decoding of the collided ACK transmissions. Our coding method does
not require precise detection of signal energy and thus there is no
modification needed for the physical-layer modulation. In
particular, each receiver assigns a unique bitstream pattern to
embed in its ACK packet. Different superimpositions of these bitstream
patterns result in different decoded bitstream, which enables the
sender to deduce the presence of individual ACK transmissions involved
in the collision. In this way, there is no need for each receiver to
transmit its ACK in different time slots.

\section{Proposed Cooperative Retransmissions Through Collisions}
\label{sect:proposed} In this section, we will show how these
interference-embracing techniques can be used in a common scenario
of two interfering pairs of sender-receiver communicating in a lossy
environment. We will use the case of two interfering WLAN APs as an
example to illustrate our idea, although it can be applied to other
wireless network scenarios as well.

\subsection{Basic Idea}
Consider the two pairs, $AP_1 \sim R_1$ and $AP_2 \sim R_2$ in
Fig.~\ref{fig:network}, in a lossy wireless network, where both
receivers are within the transmission range of the two APs. Let us
assume that $AP_1$ wishes to transmit a packet $c_1$ to $R_1$ and
$AP_2$ wishes to transmit a packet $c_2$ to $R_2$. Suppose that
after the transmission packet $c_1$ is not heard by $R_1$ but
overheard by $R_2$, while packet $c_2$ is not heard by $R_2$ but
overheard by $R_1$ due to the broadcast nature of wireless
transmission (also known as opportunistic listening~\cite{Katti07}).
In this case, rather than retransmitting each of the two lost
packets in different time slots to avoid interference, it is
possible that both $AP_1$ and $AP_2$ retransmit their packet $c_1$
and $c_2$ simultaneously, which can be decoded by the two receivers
using ANC as each of them already has one known packet. In this way,
we can improve the retransmission efficiency by reducing one
retransmission.

\subsection{Protocol Design}
In the practical scenario of two interfering APs shown in
Fig.~\ref{fig:network}, there are typically multiple receivers
associated with each AP. For receivers located in non-interference
regions, their transmission and retransmission follow the standard
IEEE 802.11 protocol. Only for receivers located in the interference
region, the retransmission is carried out using both the proposed
cooperative collision and the conventional ARQ.

In order to enjoy the proposed cooperative retransmission, two
receivers belonging to different APs in the interference region need
to be paired up. In particular, each receiver station first connects
to an AP. After the establishment of the $AP_i \sim R_i$ connection,
the receiver then detects whether it is in the interference region
by overhearing transmission from another AP, $AP_j$. If it is in the
interference region, it then broadcasts its availability to pair-up
with receiver $R_j$ connected to $AP_j$ and located inside the
interference region. If receiver $R_j$ is available, it accepts the
pairing invitation. After that, both $R_i$ and $R_j$ broadcast their
pairing information to the APs. Once paired, both $R_i$ and $R_j$
can acknowledge packets destined for anyone of them and no third
receiver is allowed to participate in acknowledging packets destined
for either $R_i$ or $R_j$.

\begin{figure}
\begin{center}
\includegraphics[width = 0.5\textwidth]{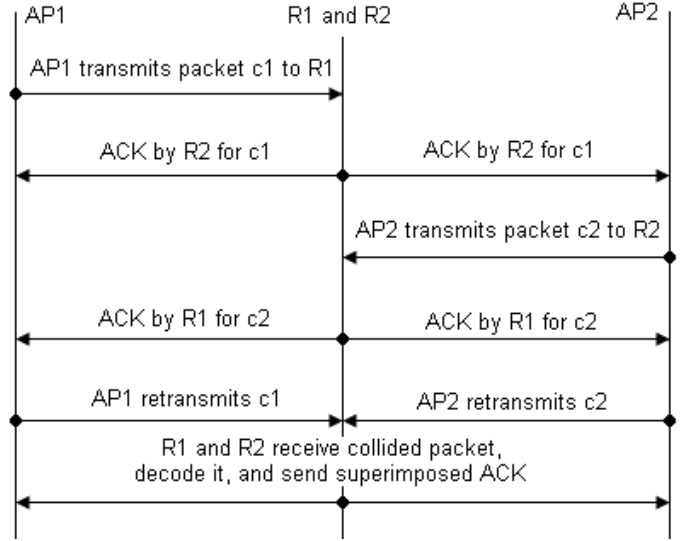}
\end{center}
\caption{Example of the handshake procedure for cooperative
retransmission.} \label{handshake}
\end{figure}

Suppose we have established the connections of $AP_1 \sim R_1$ and
$AP_2 \sim R_2$ and the pair-up of $R_1 \sim R_2$ as shown in
Fig.~\ref{fig:network}. Initially, both APs will transmit and
retransmit packets using 802.11 MAC protocol and both receivers will
reply with an ACK embedded with the collision codes~\cite{Foh10} for
every packet they hear and destined to anyone of them. If both
receivers hear the same packet, they will transmit their ACKs
simultaneously and the APs uses the aforementioned
technique~\cite{Foh10} to decoded the superimposed ACK. If $AP_1$
only detects an ACK from $R_2$ for a packet $c_1$ destined for
$R_1$, it defers the retransmission until it finds an opportunity
for cooperative retransmission. Because of the broadcasting nature
of ACK transmission, $AP_2$ is aware that $AP_1$ has deferred a
retransmission. When $AP_2$ only detects an ACK from $R_1$ for a
packet $c_2$ destined for $R_2$, $AP_2$ is then available to
participate in cooperative retransmission. Since both APs are aware
of each other's deferred retransmission status, they then
simultaneously retransmit their corresponding packets, which results
in a collision. Once the receivers successfully decode the collided
packet using ANC, they will send superimposed ACK immediately.
Figure~\ref{handshake} illustrates the handshake procedure for the
cooperative retransmission.

\section{Performance Analysis}\label{sect:analysis}
So far, we only show that there is a possibility that the
interference-embracing techniques can be utilized to improve the
retransmission efficiency in the scenario of two interfering
sender-receiver pairs. In this section, we mathematically analyze
the probability and corresponding performance gain.

\subsection{System Model} \label{sect:model}
Let $d_{AP}$ denote the distance between the two APs, and $r_t$
denote the transmission range of each AP with both APs transmitting
at the same transmission power and the same transmission rate.
Consider that the interfering APs are overlapped such that $d_{AP}<
2r_t$. Each AP associates with $N$ client stations, which are
uniformly distributed within the transmission range of the AP. Of
interest to us are the receivers located inside the interference
region. Consider the two pairs $AP_1 \sim R_1$ and $AP_2 \sim R_2$
shown in Fig.~\ref{fig:network}. Average packet loss probability
$p_{ij}$ for transmissions from $AP_i$ to receiver $R_j$ follows an
independent Bernoulli packet loss model~\cite{Salyers08}, where
$\{i,j\}\in\{1,2\}$. Packet batch size for transmissions from
$AP_i$ to $R_i$ is denoted as $B_i$. We assume that $B_1$=$B_2$=$B$.
For multimedia applications such as video streaming, $B$ is usually
a large value.

\subsection{Retransmission Efficiency}
We use ARQ as the benchmark for performance comparison. It is well
known that the average number of retransmissions needed for
recovering a lost packet follows the geometric distribution. Thus,
the average total number of retransmissions needed for both $AP_1$
and $AP_2$ to successfully deliver $B$ packets is
\begin{equation}\label{eq:ARQ}
N_{ARQ} = \sum_{i=1}^{2} \frac{B\cdot p_{ii}}{(1-p_{ii})}.
\end{equation}

In our proposed scheme, each AP builds up packet reception status
for every packet it transmits. Because of the broadcasting nature
of superimposed ACK, $AP_i$ is aware of the reception status of
both $R_i$ and $R_j$. Therefore, any transmitted packet
will have four reception states shown in
Table~\ref{table:charateristics}.

\begin{table}
\caption{Packet reception status for a packet transmitted by
$AP_i$.} \label{table:charateristics}
\begin{center}
\begin{tabular}{|c|c|c|c|}
\hline State & Definition & Probability \\
\hline 1     & received by $R_i$ but not by $R_j$   & $p_{ij}(1-p_{ii})$ \\
\hline 2     & received by $R_j$ but not by $R_i$   & $p_{ii}(1-p_{ij})$ \\
\hline 3     & received by both $R_i$ and $R_j$     & $(1-p_{ii})(1-p_{ij})$ \\
\hline 4     & not received by both $R_i$ and $R_j$ & $p_{ii}p_{ij}$ \\
\hline
  \end{tabular}
\end{center}
\end{table}

Since the packets transmitted by $AP_i$ is destined to $R_i$, both
state 1 and state 3 in Table~\ref{table:charateristics} are
considered instances of successful reception cases. For state 4, where the packet
is not received by both receivers, $AP_i$ would repeatedly
retransmit the packet in traditional ARQ fashion until the
retransmission falls into any one of the first 3 states. The total number
of retransmissions needed for changing the packet reception status for all packets
in state 4 to one of the first 3 states follows geometric distribution with
average loss probability of $p_{i1}p_{i2}$. Thus, the
total number of retransmissions needed for packets in state 4 by both the senders is calculated as
\begin{equation}\label{eq:state4}
N_{CR-S4} = \sum_{i=1}^{2} \frac{B \cdot
p_{i1}p_{i2}}{1-p_{i1}p_{i2}}.
\end{equation}

State 2 in Table~\ref{table:charateristics} is the case for
cooperative retransmission. Suppose $AP_1$ and $AP_2$ are now
simultaneously retransmitting $c_1$ and $c_2$ to $R_1$ and $R_2$,
respectively. There are two states for the reception of $c_1\odot
c_2$ at each receiver, as shown in Table~\ref{table:collision}. Note
that only when both $c_1$ and $c_2$ reach $R_i$ successfully, the
reception of the collided packet at $R_i$ is considered as a
success. This is because any corruption in one of the packets will
cause the collided packet undecodable by using ANC.
\begin{table}
\caption{Packet reception states of cooperative retransmission at
$R_i$.} \label{table:collision}
\begin{center}
\begin{tabular}{|c|c|c|c|}
\hline State & Definition & Probability \\
\hline $S_a$     &successfully receive $c_1\odot c_2$
&$(1-p_{ii})(1-p_{ji})$ \\
\hline $S_b$    &$c_1\odot c_2$ is corrupted.
&$1-(1-p_{ii})(1-p_{ji})$
\\ \hline
  \end{tabular}
\end{center}
\end{table}

Therefore, the total number of retransmissions needed for state 2
can be derived as
\begin{equation}\label{eq:state2}
N_{CR-S2} = \frac{B \cdot P_{S2, i}}{(1-p_{ii})(1-p_{ji})}
\end{equation}
where $P_{S2, i}$, the probability that a transmitted packet by
$AP_i$ is in state 2, is given by
\begin{equation}\label{eq3}
P_{S2, i} = p_{ii}(1-p_{ij})\left\{1 +
\sum_{n=1}^\infty(p_{ii}p_{ij})^n\right\},
\end{equation}
which takes into consideration additional packets falling in state 2
after retransmission of packets in state 4. Note that unlike
\eqref{eq:ARQ} and \eqref{eq:state4}, there is no summation sign in
~\eqref{eq:state2}. This is because of the collision based
cooperative retransmission, where the retransmissions for one
receiver can always be piggyback in the retransmissions for another
receiver.

It is reasonable to assume that both $AP_1$ and $AP_2$ can always
find `partner packets' in cooperative retransmission for
multimedia applications such as video streaming, which typically
have large $B$ values. In practice, if there is no `partner
packets', those lost packets are just retransmitted in the
traditional way.

Finally, we compute the total number of retransmissions needed for
our proposed cooperative retransmission as
\begin{equation}
N_{CR} = N_{CR-S4} + N_{CR-S2}.
\end{equation}
Assuming that $p_{11}=p_{12}=p_{21}=p_{22}=p$, we derive the
retransmission gain against ARQ as
\begin{equation}
G_r=\frac{N_{ARQ}}{N_{CR}}=\frac{2(1-p^2)}{2p(1-p)+1},
\end{equation}
which gives a theoretical retransmission gain of $2>G_r>1$ for
$0<p<1/2$.

We now derive the total gain for the entire network, where each AP
is associated with $N$ uniformly distributed receivers. According
to the system model in Section~\ref{sect:model} and the geometry
relationships shown in Fig.~\ref{fig:network}, we can derive the
overlapped area as
\begin{equation}
A = 2r_t^{2}\left(\arccos \frac{d_{AP}}{2r_t}\right) -
d_{AP}\sqrt{r_t^{2}-\frac{d_{AP}^{2}}{4}}.
\end{equation}
It is clear that the total network gain depends on the number of
receiver pairs located in the overlapped area, which is $N_A = N
\cdot \frac{A}{\pi r_t^2}.$ Therefore, the total retransmission
gain with respect to all receivers in the network is given as
\begin{equation}
G_N=\frac{N \cdot N_{ARQ}}{N_A \cdot N_{CR} + (N-N_A) \cdot
N_{ARQ}}.
\end{equation}

\section{Simulation Results}\label{sect:simu}
Packet decoding using ANC has been successfully demonstrated on a
test bed in~\cite{Katti07}. Therefore we can confidently assume
that ANC is a practically applicable technique. For the proposed
collision based cooperative retransmission, we construct a C++
discrete-time simulator with the system model described in
Section~\ref{sect:model}. For simplicity, we assume the network
environment for the two APs are homogeneous and symmetric, e.g.
same packet loss rate and distance between  $AP_1 \sim R_1$ and
$AP_2 \sim R_2$

\begin{figure}
\begin{center}
\includegraphics[width = 0.5\textwidth]{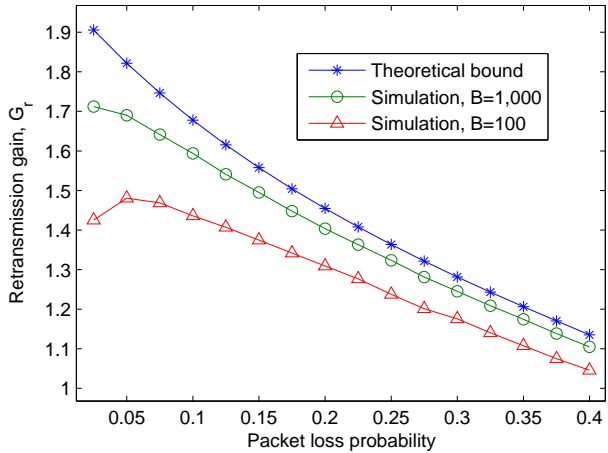}
\end{center}
\caption{Retransmission Gain $G_r$ under different packet loss
rates.} \label{fg:gain_plr}
\end{figure}

Fig.~\ref{fg:gain_plr} shows the retransmission gain $G_r$ under
different packet loss probabilities. We can see that the simulation
results with $B=1000$ matches the theoretical results well. The
relatively large difference at low packet loss rates is due to
the unavailability of `partner packet' for the cooperative
retransmission.

Compared with the results with $B=1000$ and $B=100$, we can see that
the difference between the theoretical gain and the simulation gain
becomes smaller with the increase of the batch size. This is because
a larger batch size leads to more cooperative collision coding
opportunities, which is consistent with the assumptions we made in
the theoretical analysis. On the other hand, for the case of small
batch size, the problem of no `partner packet' becomes more severe.

\begin{figure}
\begin{center}
\includegraphics[width = 0.5\textwidth]{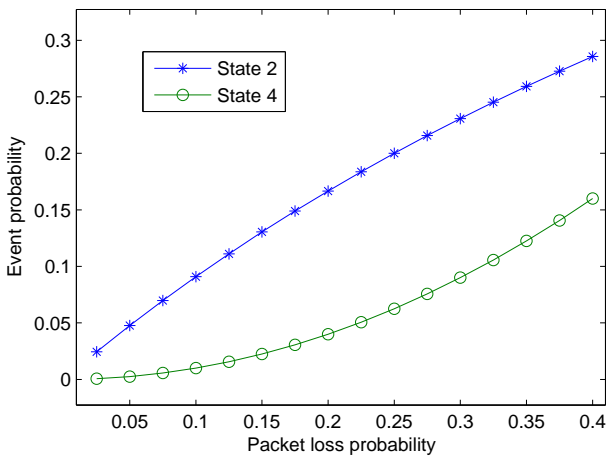}
\end{center}
\caption{Probabilities of state 2 and state 4. } \label{fig2}
\end{figure}

It can also be seen from Fig.~\ref{fg:gain_plr} that the
retransmission gain is reduced with the increase of packet loss
rate. There are two main reasons for this. First, large packet
loss rate reduces the probability of successful reception of the
collided packets as shown in Table~\ref{table:collision}. Second,
with the increase of packet loss rate, the probability for state 4
becomes significant (see Fig.~\ref{fig2}), where the traditional
ARQ based retransmission is used, and thus it reduces the gain
from the cooperative retransmission.

\begin{figure}
\begin{center}
\includegraphics[width = 0.5\textwidth]{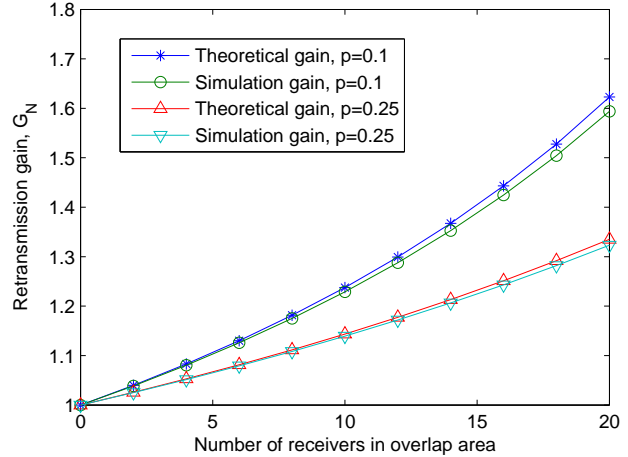}
\end{center}
\caption{Network retransmission gain $G_N$ with $N=10$ and
$B=1000$.} \label{fig3}
\end{figure}

Fig.~\ref{fig3} shows the network retransmission gain as the
overlap area increases, for which the distance between the APs
decreases. Each AP is associated with 10 uniformly distributed
receivers. With the increase of the overlapped area, more
receivers are located inside the overlapped region, where there
are more pairs for the cooperative retransmission. As expected,
from Fig.~\ref{fig3}, we can see that the network gain increases
with the increased number of receivers located in the overlapped
area.

\section{Conclusion}\label{sect:conclude}
In this paper, we have successfully applied the existing interference-embracing
techniques in the scenario of two interfering WLAN APs, which are simulcasting
bulky data to their associated individual stations in a lossy environment. In
particular, we have proposed a collision based cooperative retransmission
scheme. Our major contribution lies in the protocol design which well combines
different interference-embracing techniques to solve the retransmission
problem. We have also analyzed the performance gain of our proposed cooperative
retransmission against the traditional ARQ scheme. Both theoretical analysis
and simulations show that our proposed collision based retransmission method is
able to reduce the number of retransmission of ARQ by up to 50\%.

Although we focus on the scenario of two interfering WLAN APs, the
proposed collision based cooperative retransmission scheme can be
applied to any two interfering pairs of `sender-receiver', which
is quite common in WLANs, wireless mesh networks and wireless
sensor networks. Our future work will be to extend the proposed
scheme to more general scenarios such as a mixture of simulcasting
and multicasting receivers and heterogeneous receivers.

\bibliographystyle{IEEEtran}
\bibliography{IEEEabrv,paper}

\begin{thebibliography}{1}
\providecommand{\url}[1]{#1}
\csname url@samestyle\endcsname
\providecommand{\newblock}{\relax}
\providecommand{\bibinfo}[2]{#2}
\providecommand{\BIBentrySTDinterwordspacing}{\spaceskip=0pt\relax}
\providecommand{\BIBentryALTinterwordstretchfactor}{4}
\providecommand{\BIBentryALTinterwordspacing}{\spaceskip=\fontdimen2\font plus
\BIBentryALTinterwordstretchfactor\fontdimen3\font minus
  \fontdimen4\font\relax}
\providecommand{\BIBforeignlanguage}[2]{{%
\expandafter\ifx\csname l@#1\endcsname\relax
\typeout{** WARNING: IEEEtran.bst: No hyphenation pattern has been}%
\typeout{** loaded for the language `#1'. Using the pattern for}%
\typeout{** the default language instead.}%
\else
\language=\csname l@#1\endcsname
\fi
#2}}
\providecommand{\BIBdecl}{\relax}
\BIBdecl

\bibitem{Drieberg09}
M.~Drieberg, F.-C. Zheng, R.~Ahmad, and M.~Fitch, ``Impact of interference on
  throughput in dense {WLAN}s with multiple {AP}s,'' in \emph{IEEE PIMRC},
  Tokyo, Japan, Sept 2009.

\bibitem{Zhang06}
S.~Zhang, S.~C. Liew, , and P.~P. Lam, ``Hot topic: physical-layer network
  coding,'' in \emph{ACM MobiCom}, Los Angeles, USA, September 2006.

\bibitem{Katti07}
S.~Katti, S.~Gollakota, and D.~Katabi, ``Embracing wireless interference:
  Analog network coding,'' in \emph{ACM SIGCOMM}, Kyoto, Japan, 2007.

\bibitem{Foh10}
C.~H. Foh, J.~Cai, and J.~Qureshi, ``Collision codes: Decoding superimposed
  {BPSK} modulated wireless transmissions,'' in \emph{IEEE CCNC}, Las Vegas,
  USA, January 2010.

\bibitem{Durvey07}
M.~Durvy, C.~Fragouli, and P.~Thiran, ``Towards reliable broadcasting using
  {ACK}s,'' in \emph{IEEE ISIT}, Nice, France, 2007.

\bibitem{Akella05}
A.~Akella, G.~Judd, S.~Seshan, and P.~Steenkiste, ``Self management in chaotic
  wireless deployments,'' in \emph{ACM MobiCom}, Cologne, Germany, 2005.

\bibitem{Hamkins00}
J.~Hamkins, ``An analytic technique to separate cochannel fm signals,''
  \emph{IEEE Transactions on Communications}, vol.~48, no.~11, pp. 2980--2989,
  April 2000.

\bibitem{Salyers08}
D.~C. Salyers, A.~Striegel, and C.~Poellabauer, ``Wireless reliability:
  Rethinking 802.11 packet loss,'' in \emph{IEEE WoWMoM}, Newport Beach, USA,
  2008.

\end{thebibliography}

\end{document}